\documentclass[twocolumn,english,aps,prl,floatfix,showpacs]{revtex4}
\usepackage[T1]{fontenc}
\usepackage[latin1]{inputenc}

\makeatletter

\usepackage{babel}

\usepackage{babel}
\makeatother

\begin{document}

\title{ Gauge Invariance and Spinon-Dopon
Confinement in the $t-J$ Model:\\
Implications for Fermi Surface Reconstruction in the Cuprates}

\author{Alvaro Ferraz$^{1}$, Evgeny Kochetov$^{1,2}$}

\affiliation{$^{1}$International Institute of Physics - UFRN,
Department of Experimental and Theoretical Physics - UFRN, Natal, Brazil}
\affiliation{$^{2}$Laboratory of Theoretical Physics, Joint
Institute for Nuclear Research, 141980 Dubna, Russia}


\begin{abstract}
We discuss the application of
the two-band spin-dopon representation of the $t-J$ model to address the issue of the Fermi surface reconstruction observed in the cuprates.
We show that the electron no double occupancy (NDO) constraint plays a key role in this formulation.
In particular, the auxiliary lattice spin and itinerant dopon degrees of freedom of the spin-dopon formulation
of the $t-J$ model are shown to be confined
in the emergent $U(1)$ gauge theory generated by the NDO constraint.
This constraint is enforced by the requirement of an infinitely large spin-dopon coupling.
As a result, the $t-J$ model is equivalent to a Kondo-Heisenberg lattice model
of itinerant dopons and
localized lattice spins at infinite Kondo coupling at all dopings. We show that
mean-field treatment of the large vs small Fermi surface
crossing in the cuprates which leaves out the
NDO constraint, leads to inconsistencies and it is automatically excluded form the $t-J$ model framework.

\end{abstract}

\maketitle

\section{introduction}

The observation of quantum oscillations in the lightly hole-doped cuprates \cite{DL} is an important breakthrough
since it indicates that coherent electronic quasiparticles may exist even in the pseudogap (PG) regime.
The PG state does not exhibit a large Fermi surface (FS) enclosing the total number
of charged carriers. Instead, the FS consists of small pockets with a total area
proportional to the dopant density $x$, rather than the $1+x$ which is expected for conventional Fermi liquids (FL's).
A possible theoretical justification for this phenomenon might be the occurrence of a simultaneous setting
of a new long-range
order together with the PG phase. \cite{morr}
The resulting breaking of translational symmetry would
cut the large FS into small pieces but the Luttinger's theorem (LT) would still hold.
However, the existence or not of such translational symmetry breaking is still debatable to this date.
Moreover, even in case this
symmetry breaking is verified, the LT might still
be violated due to the proximity to the antiferromagnetic (AF) Mott insulator transition.
We thus cannot rule out the possibility that the new metallic PG state may indeed
violate the traditional LT.
As a result, the PG state truly qualifies as
a non-Fermi liquid (NFL) state which violates the LT.
If this is indeed the case, the small Fermi pockets
could then be accounted for even without any symmetry breaking.

An instructive example of such a new metallic state is provided by the so-called fractionalized Fermi liquid  (FL$^*$)
which exhibits small pockets
similar to what is observed in an AF metal, and at the same time keeps the translational symmetry intact.
Such a state manifests itself in the context of the Kondo-Heisenberg lattice model which describes
localized Heisenberg lattice spin moments coupled to a conduction band of itinerant electrons \cite{2.1,2.2}:
\begin{eqnarray}
H_{K-H}&=&\sum_{ij}t_{ij}c^{\dagger}_{i\sigma}c_{j\sigma}+J_K\sum_{i}\vec S_i
c^{\dagger}_{i\sigma}\vec\tau_{\sigma\sigma'}c_{i\sigma'}\nonumber\\
&+&J_H\sum_{ij}\vec S_i\vec S_{j}.
\label{kh}\end{eqnarray}
Here the $c_{i\sigma}$'s represent the conduction electrons and the $\vec S_i$'s are the spin local moments on
square lattice sites, with the summation over repeated spin indices $\sigma$ being implicit.
A fermionic "slave-particle" representation
of the local moments is:
\begin{equation}
\vec S_i=f^{\dagger}_{i\sigma}\vec \tau_{\sigma\sigma'}f_{i\sigma'}
\label{fs}\end{equation}
The $f_{i\sigma}$ describes
a spinful fermion destruction operator at site $i$ and the $\vec\tau$'s are Pauli matrices.

In the regime in which the Kondo coupling $J_K$ is much greater than the Heisenberg exchange $J_H$, the localized spin $f$
moments
and the spin of the conduction $c$ electrons are locked into the singlet state:
\begin{equation}
 \frac{1}{\sqrt{2}}(|\Uparrow\rangle_f|\downarrow\rangle_{c,FS}-|\Downarrow\rangle_f|\uparrow
\rangle_{c,FS}),
\label{0}
\end{equation}
where $|\sigma\rangle_f$ represents the localized spins and
$|\sigma\rangle_{c,FS}$ is a linear superposition of the conduction-electron states near the FS.  \cite{si}
As a result of this entanglement, the local moment is readily converted into a Kondo resonance in the excitation spectrum.
The conduction electrons  and the Kondo resonances are then hybridized and together they produce a conventional FL state
with a FS
enclosing the traditional Luttinger volume, which in view of that, counts the density of both $f$ and $c$ electrons.
If the conduction band is filled with $x$ electrons per unit cell, this large FS encloses $1+x$
electrons per site.

In the opposite parameter regime, $J_H\gg J_K,$  a non-FL phase can show up instead, provided
the Heisenberg $f-f$ coupling is sufficiently frustrated. As a result, the localized spins are melted into
a quantum spin liquid.
When this phase is stable, it quenches the Kondo effect. The $c$ electrons
are effectively decoupled from the $f$ spins and they are then solely responsible for
a small FS with a volume determined
only by the density of the $c$ electrons.
This violates the traditional Luttinger count and the resulting theory describes a FL$^*$ metal.
Such a small FS can be associated formally with a modified LT to take into account
$Z_2$ topological excitations ("visons") of the fractionalized spin
liquid ground state.\cite{2.1}

The idea of the Kondo-type FL$^*$--FL transition  has recently been
carried over to treat the $t-J$ model in an attempt to describe the FS reconstruction
observed in the hole-doped cuprates.\cite{wen2,sach}  After all, recent experiments have
revealed striking similarities between the high-T$_c$ cuprates
and quasi two-dimensional
heavy fermion materials (the CeMIn$_5$ family) described by the Kondo-Heisenberg model\cite{31,32}.
In fact, a variety of
physical phenomena can be accounted for by that model, such as the
NFL behavior, the different types of both magnetic
and charge ordering as well as the
unconventional superconductivity \cite{coleman}.

In the present paper, we show that the
mean-field (MF) FL$^*$ theory of the underdoped $t-J$ model
for the underdoped cuprates
is blotted out by the electron NDO constraint.
In fact, the NDO constraint drives the theory to a strong-coupling regime
not amenable to a MF treatment.
This is a manifestation
of strong electron correlations inherent in the
physics of the underdoped cuprates.
More specifically, the NDO constraint generates a $U(1)$ gauge theory in a confining phase:
the lattice spin background and the conduction dopons are strongly coupled to the gauge field fluctuations.
As a result, the weak-coupled spin-dopon MF FL$^*$ ground state is never realized in the underdoped $t-J$ model.
Instead, the lightly doped Mott regime takes place essentially at a strong spin-dopon coupling and
one runs into inconsistencies if one tries to describe such a PG phase
without taking proper account of the NDO.

In contrast, the overdoped regime is much simpler than that since the underlying spin background is represented by
a lattice of paramagnetic spins rather than by a quantum liquid of spin singlets.
Implementing the NDO constraint in this regime results in a complete magnetic screening
of the background paramagnetic lattice spins, which are then
dissolved into the conduction sea. This leads to a FS with an enhanced volume which
recovers the traditional Luttinger counting.

\section{ spin-dopon theory}

Consider the low-energy properties of the $t-J$ model on a square lattice with
\begin{equation}
H_{t-J}=-\sum_{ij\sigma} t_{ij} \tilde{c}_{i\sigma}^{\dagger}
\tilde{c}_{j\sigma}+ J\sum_{ij} (\vec Q_i \cdot \vec Q_j -
\frac{1}{4}\tilde{n}_i\tilde{n}_j),
\label{01}\end{equation}
where $\tilde{c}_{i\sigma}=c_{i\sigma}(1-n_{i,-\sigma})$ is the Gutzwiller
projected electron operator (to avoid the on-site double
occupancy), $\vec
Q_i=\sum_{\sigma,\sigma'}\tilde{c}_{i\sigma}^{\dagger}\vec\tau_{\sigma\sigma'}\tilde{c}_{i\sigma'},
\,\vec\tau^2=3/4, $ is the electron spin operator and $\tilde
n_i=n_{i\uparrow}+n_{i\downarrow}-2n_{i\uparrow}n_{i\downarrow}$.

To establish the correspondence between the low-energy physics of the $t-J$ model and the
Kondo physics one should rewrite
the degrees of freedom of the one-band $t-J$ Hamiltonian in a two-band Kondo-Heisenberg model representation.
This can be  achieved within the recently proposed spin-dopon representation of the constrained electron
operators. \cite{wen,fkm}
In terms of the $su(2)$ spin and the fermionic dopon operators, the
projected electron operators take the form \cite{wen}
\begin{equation}
\tilde c_{i\sigma}^{\dagger}=
\frac{sign(\sigma)}{\sqrt{2}}[(1/2+sign(\sigma) S^z_i)\tilde d_{i-\sigma}-S^{\sigma}_i\tilde d_{i\sigma}],
\label{02}\end{equation}
where $sign(\sigma=\uparrow\downarrow)=\pm 1.$
Here $\tilde{d}_{i\sigma}=d_{i\sigma}(1-d_{i,-\sigma}^{\dagger}d_{i,-\sigma})$ denotes the Gutzwiller projected
dopon operator, whereas
$\vec S_i^{\sigma}$ denotes the spin-raising (-lowering) operator $S_i^{+}$ $(S_i^{-})$
for $\sigma=\uparrow(\downarrow)$.
In this framework, the holes are doped carriers in the half-filled Mott insulator which can otherwise be described
exclusively in  terms of spin variables.

To accommodate these new operators one obviously needs to enlarge
the original onsite Hilbert space of quantum states. This enlarged
space is characterized by the state vectors $|\sigma a\rangle$, with
$\sigma=\Uparrow,\Downarrow$ labeling the spin projection of the
lattice spins and with $a=0,\uparrow,\downarrow$, labeling the dopon
states (the dopon double occupancy is not allowed). In this way, the on-site enlarged
Hilbert space becomes
\begin{equation}
{\cal H}^{enl}_i=\{|\Uparrow 0\rangle_i,|\Downarrow
0\rangle_i,|\Uparrow \downarrow\rangle_i, |\Downarrow
\uparrow\rangle_i,|\Uparrow \uparrow\rangle_i,|\Downarrow
\downarrow\rangle_i\},
\label{en}\end{equation}
while in the original Hilbert space we can either have one electron
with spin $\sigma=\uparrow,\downarrow$ or a vacancy:
\begin{equation} {\cal H}_i =\{|\uparrow \rangle_i,|\downarrow
\rangle_i,|0\rangle_i\}\label{3v},
\end{equation}
The following mapping between the two spaces is then defined:
\begin{equation}
|\uparrow \rangle_i \leftrightarrow |\Uparrow 0\rangle_i, \quad
|\downarrow \rangle_i \leftrightarrow |\Downarrow 0\rangle_i,
\label{spin}\end{equation}
\begin{equation} |0 \rangle_i \leftrightarrow \frac{|\Uparrow \downarrow\rangle_i
- |\Downarrow \uparrow\rangle_i}{\sqrt{2}}\label{vacancy}.
\end{equation}
The remaining triplet states $\left(|\Uparrow
\downarrow\rangle_i + |\Downarrow
\uparrow\rangle_i\right)/\sqrt{2}$, $|\Uparrow \uparrow\rangle_i$,
$|\Downarrow \downarrow\rangle_i$ are unphysical and should
therefore be removed from actual calculations. In this mapping, a
vacancy corresponds to an onsite singlet state of a
lattice spin and a dopon. The vacancy is a spin singlet entity which carries a unit charge $e$
when compared to the remaining sites.

To avoid a possible confusion, the following remark
is in order at this stage. Physically, one-hole doping corresponds to a removal of one electron, leaving
behind an
empty site, which carries a unit charge $e$ when compared to the
remaining sites. This is nothing more than a vacancy which is a dopon-spin singlet with a charge $e$.
The total number of vacancies is then
exactly equal to the total number of dopons.\cite{pfk} A hole by definition is a spin-$1/2$ object
with a charge $e$. A doped hole is then this vacancy which carries an extra spin $1/2$ spread over
the surrounding spin background. The physical hole is thus an
extended nonlocal object. 
In the doped Mott insulator the term "hole" is
often used with a different meaning. The dopons and the lattice spins are just
auxiliary gauge-dependent entities, while the hole is physical and
gauge-independent object.

Such a hole
appears as a string-like object with
much in common with the hole doped in an AF
ordered lattice introduced earlier in \cite{bulaevskii}.
This doped-hole concept was
developed further to derive an effective
single-hole gauge invariant AF action.\cite{siggia}

The
original $t-J$ Hamiltonian (\ref{01}) written in terms of the constrained operators (\ref{02})
vanishes when it acts on any of the unphysical states. Consequently,
it automatically decouples the physical and unphysical states
in the enlarged Hilbert space.\cite{wen}
Unfortunately, the $t-J$ Hamiltonian (\ref{01}) given directly in terms of the
constrained electron operators is very difficult to deal with.
This is due to the fact that
the algebra of the constrained electron operators is much more involved
than the related algebra for conventional fermion and spin operators.

To simplify the problem, one usually relies on a MF approximation.
However, some extra care needs to be exercised in this case. A
MF approximation results in a MF Hamiltonian that can no longer be written
solely in terms of the unique combinations given by (\ref{02}). As a result,
the faithful spin-dopon representation of the $t-J$ Hamiltonian is immediately destroyed by that.
Within such a MF approach, the
operators $\vec S_i$ and
$\tilde d_{i\sigma}$ act in the whole enlarged Hilbert space mixing up thereby the physical and
unphysical states. As a result, the unphysical states reappear in the theory
in an uncontrolled way.

This is precisely the case with the MF treatment exposed in Refs.\cite{wen2,sach}.
In Ref.\cite{wen2}, the MF decoupling is carried out by the spin-singlet order parameter,
\begin{equation}
\Delta_{ff}=\langle f_{i\uparrow}f_{j\downarrow}-f_{i\downarrow}f_{j\uparrow}\rangle,
\label{op1}\end{equation}
where the fermionic spinons $f_{i\sigma}$ are defined  through Eq.({\ref{fs}). This parameter is used to represent
a liquid of spin singlets at MF level.\cite{w3}
In contrast, the pairing amplitude between conduction dopons and localized spins
\begin{equation}
\Delta_{df}=\langle f_{i\downarrow}d_{i\uparrow}- f_{i\uparrow}d_{i\downarrow}\rangle
\label{op2}\end{equation}
describes the condensation of Kondo (or Zhang-Rice) spin singlets. Accordingly its nonvanishing value implies that
the localized spins contribute to the Fermi surface volume.
The underdoped FL$^*$ metallic phase is fixed by the choice $\Delta_{ff}\neq 0,\,\Delta_{df}=0$, whereas
the overdoped regime is imaged on a conventional heavy FL phase. This phase is supposed to set in under
the assumption that $\Delta_{ff}= 0$ and $\Delta_{df}\neq0$. However the precise location of the emergent small Fermi
pockets in the underdoped phase has not been determined that way.

In Ref.\cite{sach}, only the  background spin-singlet
order parameter (\ref{op1}) is used to describe the spin-liquid ground state. The fermionic amplitudes are now
replaced
by bosonic modes representing Schwinger bosons.
Since the $Z_2$ bosonic spin modes are gapped in the spin-liquid phase, they can be formally integrated out. This is done
perturbatively, by expanding the effective action in the bosonic MF propagator.
In case the emergent effective low-energy action does indeed exist (if we assume that this series converges),
the proposed theory describes the
fractionalized spin liquid weakly coupled to the conduction dopons. This is essentially the FL$^*$
hypothesis discussed above
in the context  of the Kondo-Heisenberg model. Such an approach
does not break the translational symmetry and results in a small hole-like FS for the underdoped phase
around $(\pi/2,\pi/2)$ and the symmetry related points in the Brillouin zone.

However, the approach advocated in Ref.\cite{sach}
cannot be used to describe a conventional FL state with a large
FS at large doping. Within that MF scheme, the large FS can be accounted for
provided the bosons are replaced by Schwinger fermions.
In the cuprates, one should expect yet another reason for such a reconstruction
in view of the depletion of the mobile carriers.
The FS transition discussed in Ref.\cite{wen2} is determined
by the variation of the strength of the effective spin-dopon
coupling rather than by a change in the doping level.

The discussed MF approaches also imply
that the FL$^*$ ground state
is in fact constituted of conduction dopons nearly decoupled from
the lattice spins. In Ref.\cite{wen2}, this is explicitly enforced by setting
$\Delta_{df}=0$ in the underdoped phase,
whereas in Ref.\cite{sach},  this is implicit in the assumption that
the perturbative expansion of the spin-dopon effective action
converges. Although this appears to be a more accurate treatment of the underdoped phase
than simply setting $\Delta_{df}=0,$
this convergence necessarily implies that the spinon gap is the largest energy scale in the problem
and this is not the case, in the infinite Kondo coupling regime.

A given  ground-state MF theory is only reliable if it is stable against
quantum fluctuations that manifest themselves beyond such zeroth MF order. In the
standard slave-particle theories of strongly correlated electrons, those fluctuations are due to an emergent local
$U(1)$ gauge field that takes care of the redundancy of the associated  slave-particle representations.
If that gauge field is in a confining phase,
the bare slave-particle excitations are strongly coupled to each other. Accordingly,
in this phase, all true physical excitations must be
gauge singlets.

The gauge redundancy specific for the
spin-dopon representation (\ref{02}) should not be ignored (see in this respect Refs.\cite{wen2,sach}).
Such a neglect implies that the dopons and lattice spins carry no emergent $U(1)$ gauge charges.
This is indeed a necessary condition
to justify a FL$^*$ phase characterized by  weakly coupled  well-defined dopon and lattice spin excitations.
We show however that the spin-dopon theory is inevitably a strongly coupled $U(1)$
gauge theory like any slave-particle representation of strongly correlated electrons should be.\cite{nayak}
The underlying slave particles - the dopons and the lattice spins -
are in fact gauge dependent and they are not present in the physical spectrum in a confining spin-dopon phase
that describes strongly correlated electrons.

\section{Emergent $U(1)$ gauge theory}

Since the emergent $U(1)$ gauge theory plays an essential role in the spin-dopon
formulation of the $t-J$ model, we provide below a brief account of its origin.

To begin with, there is an obvious redundancy in the spin-dopon decomposition of the constrained electron operator given
by Eq.(\ref{02}), since
the r.h.s. of this equation exhibits two extra
degrees of freedom. This redundancy is taken care of by the emergent local $U(1)$ gauge symmetry generated
by the NDO constraint.

In terms of the projected electron operators, that constraint reads
\begin{equation}
\sum_{\sigma}\tilde{c}_{i\sigma}^{\dagger}\tilde{c}_{i\sigma}+\tilde{c}_{i\sigma}\tilde{c}_{i\sigma}^{\dagger}=I.
\label{ndoc}\end{equation}
It simply  states that there are no on-site doubly occupied electron states. What is important is that the l.h.s of
Eq.(\ref{ndoc}) commutes with the constrained electron operators and, hence, with the $t-J$ Hamiltonian (\ref{01})
as well.
In the spin-dopon representation (\ref{02}), this constraint takes the form of a Kondo-type interaction: \cite{fku}
\begin{equation} \label{constr} \vec{S_i} \cdot
\vec{M_i}+\frac{3}{4}(\tilde{d}_{i\uparrow}^{\dagger}\tilde{d}_{i\uparrow}+
\tilde{d}_{i\downarrow}^{\dagger}\tilde{d}_{i\downarrow})=0,
\label{constr}\end{equation}
with
$$\vec M_i=\sum_{\sigma,\sigma'}\tilde{d}_{i\sigma}^{\dagger}\vec \tau_{\sigma\sigma'}\tilde{d}_{i\sigma'}.$$

This requirement excludes the unphysical spin-dopon triplet states in a self-consistent way,
since the operator
$$\Upsilon_i^{sd}:=\vec{S_i} \cdot
\vec{M_i}+\frac{3}{4}(\tilde{d}_{i\uparrow}^{\dagger}\tilde{d}_{i\uparrow}+
\tilde{d}_{i\downarrow}^{\dagger}\tilde{d}_{i\downarrow}),\quad (\Upsilon_i^{sd})^2=\Upsilon_i^{sd},$$
commutes both with $\tilde{c}_{i\sigma}$  and with
the $t-J$ Hamiltonian.

In view of this commutation, the local operator $\Upsilon_i^{sd}$
generates a $U(1)$ gauge symmetry:
$$\tilde c_{i\sigma}\to e^{i\Upsilon_i^{sd}\theta_i}\tilde c_{i\sigma}e^{-i\Upsilon_i^{sd}\theta_i}=\tilde c_{i\sigma}.$$
In contrast, the slave particles -- the dopons and lattice spins-- are not invariant under the action
of $\Upsilon_i^{sd}$ since
$$[\Upsilon_i^{sd}, \tilde d_{i\sigma}]\neq 0,\quad [\Upsilon_i^{sd}, \vec S_i]\neq 0.$$
In spite of that, all the physical observables constructed out of the slave operators, e.g., the on-site electron
spin operator $\vec Q_i,$  as well as the dopon number operator $\tilde n_i^d$
are gauge invariant quantities.\cite{pfk}
The MF Hamiltonians in \cite{wen2,sach} are gauge dependent and they do not commute
with $\Upsilon_i^{sd}$. In other words, the MF Hamiltonians act in the enlarged Hilbert space
that includes the unphysical states as well.

The origin of the emergent spin-dopon $U(1)$ gauge symmetry and
that of the traditional slave-particle representations
is one and the
same:  they are generated by the
electron NDO constraint. For example, the slave boson decomposition
of the electron operator
\begin{equation}
\tilde c_{i\sigma}=b^{\dagger}_if_{i\sigma},
\label{sf}\end{equation}
where $b_i$ is supposed to carry charge of the electron while the fermion $f_{i\sigma}$ carries the spin,
implies an electron NDO constraint in the form
\begin{equation}
\Upsilon_i^{fb}:=\sum_{\sigma}f_{i\sigma}^{\dagger}f_{i\sigma}+b_i^{\dagger}b_i=1.
\label{sfc}\end{equation}
Again, the representation (\ref{sf})
is invariant under the local $U(1)$ transformations generated by $\Upsilon_i^{fb}$,
$$\tilde c_{i\sigma}\to e^{i\Upsilon_i^{fb}\theta_i}\tilde c_{i\sigma} e^{-i\Upsilon_i^{fb}\theta_i}=\tilde c_{i\sigma},$$
which takes care of the redundancy exposed in (\ref{sf}).
However, as opposed to that, the redundant fields are not gauge invariant,
$$b_i\to e^{i\Upsilon_i^{fb}\theta_i}b_i e^{-i\Upsilon_i^{fb}\theta_i}=e^{i\theta_i}b_i,$$
$$f_{i\sigma}\to e^{i\Upsilon_i^{fb}\theta_i}f_{i\sigma} e^{-i\Upsilon_i^{fb}\theta_i}=e^{i\theta_i}f_{i\sigma},$$

Differently from the standard slave-particle representations, the spin-dopon NDO constraint does not simply reduce to
an operator identity that involves only the number operators of the redundant particles.
Since the spin-dopon representation engages the local $SU(2)$ spins along with
the projected fermion operators, the NDO constraint $\Upsilon_i^{sd}$ takes on
a more intricated form. It includes both the dopon number operator and the spin-dopon Kondo interaction.
As in the standard slave-particle descriptions,
the emergent $U(1)$
gauge field in the spin-dopon representation has no dynamics of its own
and, hence, it can be considered at infinite coupling.
Consequently, as we show next, the gauge dependent bare dopons and spins
are strongly coupled and are necessarily confined.

This confinement is in some sense similar to the flux-charge
"entanglement" observed in the fractional quantum Hall (FQH) effect.
The effective low-energy theories of FQH states are $U(1)$ Chern-Simons
(CS) gauge theories. The CS gauge field as  well has no independent
dynamics of its own: the CS coupling is a pure constraint. The only effect
of such coupling is to attach magnetic fluxes to charged particles.
Within the spirit of the  Anderson resonating-valence-bond concept of
incompressible quantum spin liquid, this tying of flux to                                                                                                                                                                                                                                                                                                                                                                                                                                                                                                                                                                                                                                                                                                                                                                                                                                                                                                                                                                                                                                         charge translates into a spin-flux one as discussed in \cite{kl,l}.

As known, the standard slave-particle theory can be explicitly reformulated as a $U(1)$ gauge theory \cite{baskaran} in its
confining phase. \cite{nayak}
This can be done in this way because the underlying NDO constraint has a very simple form: it
just fixes the total number of the on-site auxiliary particles.
In contrast, the spin-dopon NDO constraint
goes beyond that and this hinders the explicit derivation of the corresponding gauge theory for the $t-J$ model.
In spite of that, the NDO constraint in the spin-dopon representation offers a different way
to prove explicitly that the bare spins and dopon excitations are indeed strongly coupled to each other.
To see this we demonstrate below the equivalence
of the $t-J$ Hamiltonian
in the spin-dopon representation and the Kondo-Heisenberg lattice model
with an {\it infinitely} large
Kondo coupling, $J_K\to +\infty$.
As we already mentioned,
the necessary condition for the onset of the  FL$^*$ phase is
$J_H\gg J_K$. This condition is never realized in the $t-J$ model.
The infinitely strong Kondo coupling regime obviously rules out
such a possibility.

\subsection{t-J model vs Kondo-Heisenberg model}

To establish explicitly the correspondence between the infinitely coupled
Kondo-Heisenberg lattice model and the $t-J$ model,
let us notice that
the set of local constraints ${\Upsilon}_i^{sd}=0$, one
for each lattice site, is equivalent to the global condition
$\Upsilon^{sd}:=\sum_i \Upsilon_i^{sd}=0$.
This simplification holds true because the unphysical states manifest themselves as the
degenerate eigenvectors of ${\Upsilon}_i^{sd}$ with an eigenvalue 1.
Therefore, if it acts on an unphysical state, $\Upsilon^{sd}$ simply
produces the same state multiplied by a positive number. Contrary to that, acting on a
physical state, $\Upsilon^{sd}$ always gives zero.
We can therefore enforce the local constraint ${\Upsilon}_i^{sd}=0$ by adding  an extra piece
to the Hamiltonian
$$\Delta H_{\lambda}=\lambda\sum_i{\Upsilon}_i^{sd},$$ with the {\it global} parameter $\lambda$ being sent to $+\infty$.
In this way, all the unphysical states are separated from the physical
spectrum by an energy gap $\sim\lambda$.
In the limit $\lambda\to +\infty$,
they are  automatically excluded.

The
$t-J$ Hamiltonian in the spin-dopon representation then reads \cite{pfk}
$$H_{t-J}= \sum_{ij\sigma} (2t_{ij}+\frac{3\lambda}{4}\delta_{ij})\tilde{d}_{i\sigma}^{\dagger} \tilde{d}_{j\sigma}
+\lambda
\sum_{i}\vec{S_i} \cdot \vec{M_i}$$
\begin{equation}
+ J\sum_{ij} (\vec S_i \cdot \vec S_j -
\frac{1}{4})(1-\tilde{n}^d_i) (1-\tilde{n}^d_j),
\label{03}\end{equation}
with $\lambda$ being sent to $+\infty$ to ensure the selection of the physical subspace.
As we show in the Appendix A, this representation is indeed
equivalent to the standard $t-J$ model, and it reproduces the well known $1d$ exact result.

Close to half filling, where the density of doped holes
is small $x:=\langle \tilde n^d_i\rangle \ll 1$, one can also make the change $J\to \tilde J=J(1-x)^2.$
One can safely ignore the ``tilde'' sign
for the dopon operators as well, since the NDO constraint for the dopons is already taken care of by
the requirement ${\Upsilon}^{sd}_i=0.$
The spin-dopon representation of the $t-J$ Hamiltonian for
the underdoped cuprates then takes a form of the Kondo-Heisenberg lattice model, namely
\begin{equation} H_{t-J} = \sum_{ij\sigma}
T_{ij}{d}_{i\sigma}^{\dagger} {d}_{j\sigma}+
\tilde{J}\sum_{ij} (\vec S_i \cdot \vec S_j - \frac{1}{4}) +\lambda
\sum_{i}\vec{S_i} \cdot \vec{M_i}
\label{04}\end{equation}
where $T_{ij}=2t_{ij}+(3\lambda/4-\mu) \delta_{ij}$.
The global parameter $\lambda$ should be send to $+\infty$ only after the thermodynamic
limit is explicitly carried out.
Although ${\Upsilon}^{sd}_i$ no longer commutes with $H_{MF}$, the limit $\lambda\to +\infty$ still singles out
the on-site physical subspace self-consistently at any instance of time. This is precisely the case because
the operator ${\Upsilon}^{sd}_i$ has no negative
eigenvalues.

Notice that the Kondo coupling $\lambda$ is present in the dopon dispersion as well.
This ensures that the energy of the
system remains finite even when $\lambda \to \infty$. This
important renormalization of the dopon dispersion is absent in earlier
attempts to establish the Kondo and the $t-J$
model correspondence.\cite{lacroix} Note that it is precisely the
local NDO constraint that is behind such a correspondence.

It is also important to stress that
the parameter $\lambda$ cannot be absorbed in the dopon chemical potential.
To see that suppose we include $\lambda$ into $\mu$
and take the limit, $\lambda \to\infty$ . If there were no more $\lambda$-dependent terms  in
the Hamiltonian, this would result in the constraint $n^d_i=0,$ which
means that the dopon band becomes empty in this limit. However, $\lambda$ enters the Kondo term
as well. This implies instead that $3/4(n^d_i)+\vec S_i\vec M_i=0,$ which immediately brings
an occupied dopon band back to the stage.

The conventional slave-particle representations allow for a similar treatment in terms
of the gauge independent variables. For instance, one can use the slave-boson representation (\ref{sf})
of the
$t-J$ Hamiltonian free of any constraints, provided an extra term
\begin{equation}
\lambda\sum_{i}({\Upsilon}^{fb}_i-1)^2,\quad \lambda\to +\infty
\label{17}\end{equation}
is added to the Hamiltonian. It explicitly singles out the physical subspace.
It is also clear that this extra term results in an infinitely strong interaction between the slave particles.
A similar approach that involves an infinitely large coupling to fix an appropriate physical Hilbert space
was successfully used to describe the Kondo effect in metals \cite{abr} as well as
the thermodynamics of the quantum Heisenberg model \cite{lar}.

An explicit MF treatment of the Kondo-Heisenberg model (\ref{kh}) at large though finite values of the Kondo coupling
can be found in Ref.\cite{ew}. It has been established that the competition between the Kondo coupling
and the Heisenberg exchange does lead to a doping driven phase transition between states with different FS volumes.
For small $J_H$, a nonvanishing solution $\Delta_{df}\neq 0$ exists down to $x=0$. Accordingly, there is no phase transition
down to $x=0$ for small enough $J_H/t.$  If $J_H$ increases, there is an extended range of small $x$
where $\Delta_{df}=0.$
This implies that for large enough $J_H/t$, there is a crossover at some $x_c$. For $x<x_c$, the MF theory predicts
a spin liquid ($\Delta_{ff}\neq 0$) with a small FS around $(\pi,\pi)$. This disagrees with
experiment because the pockets are observed
at $(\pi/2,\pi/2)$ and other symmetry related points. This deficiency of the MF treatment
is attributed to the neglect of correlations between the localized spins and the conduction holes,
which are clearly present for large $J_K/t$. One may therefore expect that a strong coupling of
the conduction hole pocket to the AF spin fluctuations will eventually create hole pockets centered at
$(\pm\pi/2,\pm\pi/2).$

\section{Strong coupling limit}

The physical regime of the parameters
to discuss the $t-J$ model within the representation (\ref{04}) is $\lambda>>t>>J$.
A description of both large and low doping phases in the strong-coupling picture is required, which may be expected
to hold best in the limit $\lambda/t\gg 1.$
In the present Section, we show that the overdoped phase does admit a reliable description in this limit, although
for the underdoped phase
the appropriate strong-coupling theory is not yet complete.

\subsection{overdoped regime}

Let us consider first the overdoped cuprates which is expected to be described by a standard FL.
In the limit $\lambda\to\infty$, we  can employ a framework which was originally put forward to treat the
full Kondo screening regime in Ref.\cite{lacroix} (see also Refs.\cite{si,ew}).
Namely, in the limit $\lambda\to\infty$, the bare vacuum state
reads
\begin{equation}
|\Psi_0\rangle ^{overdoped} =\prod_i|0\rangle_i=2^{-N/2}\prod_i(|\Uparrow \downarrow\rangle_i
- |\Downarrow \uparrow\rangle_i).
\label{06}\end{equation}
This is a product of local Kondo (Zhang-Rice) singlets and it is the ground state of the model
for $t/\lambda=J/\lambda=0$ at $x=1.$
It then follows that
the on-site vacancy state is destroyed by the operators $d^{\dagger}_{i\sigma}$:
\begin{equation}
d^{\dagger}_{i\sigma}|0\rangle_i=0.
\label{eqs}\end{equation}
In the truncated Hilbert space, the only possible excitation above the ground state take the form \cite{lacroix}
\begin{equation}
|\sigma,0\rangle_i=\sqrt{2}sign(\sigma)\,d_{i,-\sigma}\,|0\rangle_i, \quad sign(\sigma=\Uparrow, \Downarrow) =\pm1.
\label{061}\end{equation}
A free local spin moment thus behaves as an anti-particle excitation of the $d$-field above the ground state
(\ref{06}) with a kinetic energy of order
$D\sim t$, where $D$
is the conduction dopon bandwidth. Notice that for an infinite $\lambda$, the paramagnetic susceptibility
is not given by $1/T_K$, which is zero here, but by $1/D$.  \cite{zero}
The excitations
$S^{\pm}_i|0\rangle_i$ do not appear in the theory because the states $|\Uparrow\uparrow\rangle_i$
and $|\Downarrow\downarrow\rangle_i$
have been already excluded by the NDO constraint.

Since the dopons represent holes, the local spin moment now behaves as the conduction
``electron''  with the quantum numbers, spin $1/2$ and charge $-e$ (when compared to the vacuum state).
This is a direct consequence of the
infinitely strong Kondo screening, or equivalently, of the exact resolution of the NDO constraint.
Under the assumption that the LT holds in this case, we can then conclude that the FS encloses
$1-x$ electron-like particles or, equivalently,
$(1+x)$ holes per unit cell (There are two possible states per unit cell). This is a large hole-like FS.
This phase only sets in provided the dopon hopping effectively destroys all the spin singlets when
the local AF order disappears. This marks the termination of the PG phase.
In this way, the necessary energy to break the spin singlet is roughly $J$. Since the dopon kinetic energy is
of order $2tx$,
for a hole doped Mott insulator, this only happens when
$x\ge x_c=J/2t\approx 0.15$ (for $J=t/3$).

\subsection{underdoped phase}

Next we switch to the
lightly doped regime, $x\ll 1$.
This is a more involved case, since the physics behind this phase is still unclear.
A common belief is that it is essentially determined by strong electron correlations
encoded in the NDO constraint.
To illustrate a generic difficulty that hinders a non-MF treatment in this case,
we briefly discuss a recently proposed approach \cite{ew} to
deal with
the underdoped phase seemingly beyond a MF approximation.
In that paper, the strong-coupling theory ($\lambda>>t>>J$) of the
Kondo-Heisenberg model is considered at small doping.
The corresponding $\lambda$-stabilizing term is not included in that approach, however.
A similar approach has been employed to treat
the Hubbard model in the limit $U/t>>1$ and slightly away from half-filling.
\cite{ewo}

The following remark is in order  at this stage.
The authors of Refs. \cite{ew,ewo} claim that the single-band Hubbard model as well as the $t-J$ model
can be derived
from the Kondo-Heisenberg lattice model in the limit of large Kondo coupling.
However, they provide no explicit derivation of that.
As will be argued in the Appendix A, the $t-J$ model is indeed identical to
the strongly coupled Kondo-Heisenberg lattice model.
However, this correspondence implies both the infinitely large Kondo coupling regime
and a {\it simultaneous} renormalization of the
the hopping amplitude, $t_{ij}\to T_{ij}(\lambda)$ as given
by our representation (\ref{04}).

In the region $x<<1$, the
Kondo-Heisenberg model is assumed to display short-range AF spin fluctuations.
The bare vacuum at $x=0$ is then taken to be a spin-liquid state
$|\Psi_0\rangle^{underdoped}.$\cite{ew,ewo} In contrast to the overdoped regime
where the state $|\Psi_0\rangle^{overdoped}$ is the exact eigenstate of the
strongly coupled Kondo-Heisenberg Hamiltonian,
at $x=1$,
the proposed bare vacuum state is not an eigenstate of the Hamiltonian
at $x=0$.
A precise form of this state is therefore not specified.
What is important is that this state has exactly one spin per site,
has momentum zero, and is a spin singlet. In particular, one can write it
in the form of the resonating valence bond (RVB) spin singlet:
\begin{equation}
|\Psi_0\rangle^{underdopded}\equiv |\Psi_0\rangle=|RVB\rangle\otimes|vac\rangle,
\label{07}\end{equation}
where
$$|RVB\rangle\sim\sum _{c_{\{ij\}}}\prod_{ij}c_{\{ij\}}
(|\Uparrow_i \Downarrow_j\rangle
- |\Downarrow_i \Uparrow_j\rangle),$$
and $|vac\rangle$ stands for a canonical fermionic vacuum state.
The coefficients $c_{\{ij\}}$'s are such that the resulting spin-spin correlation length is finite.
The only physical quantity which is claimed to be relevant for the calculation of the quasiparticle spectrum
is the static spin-spin correlation function,\cite{ew,ewo}
\begin{equation}
\chi_{ij}=\langle\Psi_0|\vec S_i\cdot\vec S_j|\Psi_0\rangle.
\label{chi}\end{equation}
This is considered as an input parameter.
Upon fixing in this way the spin sector of the Hilbert space, the authors proceed to
a description of the charge excitations on top of it. It is clear that such an approach displays
no dynamical
correlations between the spin and charge degrees of freedom.

In the limit of the large Kondo coupling, the charge sector comprises the on-site spin-dopon singlet
and triplet states. The triplet state corresponds to a higher energy and it is separated from the
lowest singlet state by a gap $\sim \lambda$. By an appropriate redifinition, the energy of the spin-singlet state
can be taken to be finite in the limit $\lambda\to\infty$. This limit then pushes the triplet states
out of the physical spectrum. Since it is precisely this case that has a direct relevance for the $t-J$ and Hubbard models,
we adjust  the results exposed in Refs.\cite{ew,ewo}
to that situation exclusively.

The charged quasiparticle excitations above the spin ground state can then be taken in
the form
\begin{equation}
|\vec k,\sigma\rangle = \sum_i\tilde a^{\dagger}_{i\sigma}e^{i\vec k\vec R_i}|\Psi_0\rangle.
\label{08}
\end{equation}
Here
\begin{equation}
\tilde a^{\dagger}_{i\sigma}=\tilde c_{i,-\sigma},
\label{09}
\end{equation}
where the constrained electron (Hubbard) operator $\tilde c_{i\sigma}$ is given by our Eq.(\ref{02}).
The action of the
fermionic operators $\tilde a^{\dagger}_{i,-\sigma}$
on a localized spin state $|\sigma,0\rangle_i$ produces a vacancy state, e.g.
\begin{equation}
\tilde a^{\dagger}_{i\downarrow}|\Uparrow,0\rangle_i =\frac{1}{\sqrt{2}}(|\Uparrow \downarrow\rangle_i
- |\Downarrow \uparrow\rangle_i).
\label{10}
\end{equation}
This process effectively describes the effect of hole doping in AF spin background. It appears as a vacancy
surrounded by a locally disturbed spin-liquid background.

The single-hole spin-dopon wave function (\ref{08})
describes a "dressed" hole in analogy with the many-body wave
function which was
used to describe, in the context of a spin-wave approximation,  the AF
string or the spin polaron associated with the $t-J$ model in the presence
of AF ordering. \cite{belinicher} The important distinction, in our case, is the fact that
the vacancies are now inserted in a spin-liquid background rather than in
the $N\acute{e}el$ state which was used to characterize the AF lattice.

The fermion operator $\tilde a_{i\sigma}^{\dagger}$, which transforms itself in the fundamental $SU(2)$ representation
\cite{ritt},
creates a quasiparticle with spin $\sigma$ and charge $e$. The dopon operator $d_{i\sigma}^{\dagger}$
produces the same effect when acting on the on-site canonical vacuum state $|0\rangle_i.$
At $\lambda=0$, the Kondo-Heisenberg model reduces to a gas
of noninteracting dopons decoupled from the spin background. Let now the spin-dopon
interaction $\lambda$ be adiabatically turned on towards large values.
It is then assumed that the resulting final state is
a gas of the quasiparticles (\ref{08}) weakly coupled to the same spin background.
In other words, those quasiparticles are assumed to be low-energy excitations in the physical spectrum.
This is the key assumption in Refs.\cite{ew,ewo}.
In this case,
the low-energy excitations in the quasiparticle sector take the form
\begin{equation}
E_{n_{\vec k}}=\sum_{\vec k\sigma}E_{\vec k} d^{\dagger}_{\vec k\sigma}d_{\vec k\sigma}+ \cdots,
\label{11}\end{equation}
for some $E_{\vec k}$. The omitted terms in (\ref{11})
describe weak interactions between the quasiparticles.
The crucial point is the replacement of
the constrained fermion operator $\tilde a_{\vec k\sigma}$
by the conventional unconstrained dopon operator, $d_{\vec k\sigma}$

One thus arrives at a FL state described in terms of conventional quasiparticles.
As shown in Ref.\cite{ew}, the FS in such a scheme encloses a volume proportional
to the density of the doped holes.
This does not occur as a consequence of the backfolding
of the Brillouin zone due to any kind of broken symmetry.
If one further assumes that the spin-liquid background
forms a $Z_2$ spin liquid, one  can
easily prove that the modified LT holds true in this case as well.
The derived FS is then directly associated with both dopon and
$Z_2$ gauge excitations with
a FL$^*$ state set up as advocated earlier in other MF treatments.\cite{wen2,sach}

Let us see now in what way the NDO constraint modifies
the theory discussed in \cite{ew,ewo}.
If we assume that at low doping the NDO constraint is not so important and it is relaxed,
the bilinear form
\begin{equation}
\sum_{ij\sigma}t_{ij}\tilde a^{\dagger}_{i\sigma}\tilde a_{j\sigma}
\label{12}\end{equation}
constructed out of the constrained electron operators
is replaced by the new kinetic term
\begin{equation}
\sum_{ij\sigma}t^{eff}_{ij}d^{\dagger}_{i\sigma}d_{j\sigma},
\label{12.1}\end{equation}
where the $d_{i\sigma}$'s
represent the canonical unrestricted dopon operators and $t^{eff}_{ij}$ is a certain effective
hopping amplitude
(see the Appendix C).

However, the unphysical states in this formalism are not just doubly occupied dopon states. The triplet spin-dopon states that
involve only single occupied dopon states are unphysical as well.
It is the NDO constraint that eliminates all of the unphysical states.
In spite of the fact that the operators $\tilde a_{i\sigma}$
bear the same quantum numbers as
the dopon operators, the algebra they are closed into is much more complicated than
the canonical fermionic algebra. In fact, the constrained fermion operators obey
in the configuration space the $su(2|1)$ superalgebra
commutation/anticommutation relations \cite{wiegmann} that mix up the bosonic and fermionic degrees of freedom.
In general, the bilinear form (\ref{12}) cannot be diagonalized
neither in the configuration nor in the
momentum spaces. \cite{koch} The only exceptions are the $1d$ case discussed in the Appendix A and
in the case of an exactly one hole doped into an AF spin background, the so called Nagaoka phase. \cite{nag}.

Although there is indeed
a low probability for two holes to hop on the same site in the low doping regime, relaxing the on-site
NDO constraint
drastically affects the physics at any doping level, not just at high dopings as
usually claimed elsewhere.
The operators
$\tilde a_{i\sigma}$ act in the physical Hilbert space.
However, the substitution
\begin{equation}
\tilde a_{i\sigma}\to d_{i\sigma}
\label{13}
\end{equation}
brings the unphysical triplet states back to the theory at any doping, in spite of the fact that
the strong-coupling Kondo regime is at work.

The MF relaxation of the NDO constraint modifies the underlying Hilbert space. Such a modification
results in dramatic consequences for the low-energy properties of the electron systems and this is
totally ignored by the substitution (\ref{13}).
For instance,
the approach advocated in \cite{ew,ewo} is expected to work well for $U>>t$ in the lightly doped Hubbard
model. The limit $U\to+\infty$ directly eliminates doubly occupied states, so that the resulting
Hamiltonian describes a system of strongly correlated
electrons (see Appendix B). However,
the resulting effective quasiparticle Hamiltonian given by Eq.(10) in Ref.\cite{ewo} reads
\begin{equation}
H_{U=\infty}=\sum_{ij\sigma}\tilde t_{ij}h^{\dagger}_{i\sigma}h_{j\sigma},
\label{14}\end{equation}
where $h_{i\sigma}$ is a canonical hole-like fermion operator and
\begin{equation}
\tilde t_{ij}=t_{ij}(\frac{1}{2}+2\chi_{ij}).
\label{15}\end{equation}
This is a trivial problem
that admits an exact solution in any dimensions. It is well known, however, that the $U=\infty$ Hubbard
model ( given by Eq.(\ref{1}) in the Appendix A) captures an extreme limit of the physics of strong electron correlations.
That model is certainly far from
trivial and it admits an exact solution only in $1d$.  The substitution (\ref{13}) which is the key
assumption behind such approximation obviously leaves out the essence of the physics
of the underdoped $t-J$ model, i.e., the strong electron correlations. This approach therefore reduces to a
kind of uncontrolled  MF treatment. It starts with the MF ansatz (\ref{chi}) to fix a spin-liquid
structure for the lattice spin background and proceeds by considering the dopons to be nearly decoupled
from the static spins.

Within this theory,
the positions of the hole pockets of the Hubbard model is centered at $(\pm\pi/2,\pm\pi/2).$
The pockets move to the inner side of the magnetic Brillouin zone, as the strength of the AF correlator $\chi_{ij}$
is increased.
Basically the same conclusion was reached in the MF FL$^*$ theory of the underdoped $t-J$ model.
This finding agrees with experimental data.
However, this conclusion is solely based on a choice of the input parameter
(\ref{15}). If one sets $\chi_{ij}=0$, the hole pocket moves back to $(\pi,\pi)$.
This is obviously an artifact of the MF approach rather than a true physical property
of the model.

To justify the FL$^*$ theory, the final state needs to be
adiabatically connected to the state of weakly interacting spinons and dopons.
Moreover, the stability of the MF FL$^*$ theory implies that the resulting physical quasiparticles
are just renormalized spinon and dopon excitations. However,
this is not the case for the underdoped  $t-J$ model.
Whatever small but non-zero
the doping concentration $x$ may be, the dopons
couple infinitely strongly to the lattice spins.
The true final state cannot therefore be adiabatically connected
to a state of weakly interacting dopons and spinons. (In the $1d$ case, this is explicitly demonstrated
in our Appendix A.)
The spin-dopon entanglement due to the NDO constraint
is in fact the key ingredient to discuss the underdoped phase.
To work out the relevant true low-energy spectrum one needs to resolve the NDO constraint explicitly prior
to any MF treatment.

\section{conclusion}

The spin-dopon decomposition of the constrained electron operator
is shown to be invariant under the local $U(1)$ gauge transformations generated by the local NDO constraint.
This symmetry emerges from the redundancy inherent
in the spin-dopon representation. It has been missed in earlier developments on the FS reconstruction
addressed in the framework of the spin-dopon representation.
Since the emergent gauge field is at an infinitely strong coupling,
it is necessarily a confining gauge field: the lattice spin background and the conduction dopons are always
strongly coupled to each other through the confining $U(1)$ gauge field.
In the $U(1)$ confining phase, the unphysical states are naturally excluded from the spectrum.

On the other hand, the stability of the MF  FL$^*$ ground state necessarily implies
that the $U(1)$ gauge symmetry is spontaneously broken. This contradicts a well-known assertion that
a local gauge theory can never be broken.\cite{elitzur}  Thanks to the NDO constraint there is never
a deconfining phase in which the spinons and dopons are weakly coupled to each other.
At the moment, we cannot formulate explicitly the resulting
strongly coupled $U(1)$ gauge theory of the $t-J$ model within the spin-dopon representation.
However we show explicitly that
the $t-J$ model is in fact equivalent to a Kondo-Heisenberg lattice model
of dopons and
lattice spins at infinite Kondo coupling, for all dopings.
This observation leads to the conclusion that the dopons and spinons are always confined.

MF Hamiltonians that ignore NDO are gauge dependent: they do not commute
with the local operator that enforces the constraint.
They act in the enlarged Hilbert space
that includes both physical and unphysical states.
The NDO constraint plays a key role in describing the FS crossover as a function of doping in the
$t-J$ model.
In both the overdoped as well as the underdoped phases, there is a strong entanglement
of the spin-dopon  degrees of freedom due to the NDO constraint.
The weak-coupling MF treatment of the large vs small Fermi surface
crossing in the cuprates is blotted out by the
NDO constraint that drives this model to a strong-coupling regime.

\section{Appendix A}

Here we prove that Eq.(\ref{03}) is indeed equivalent to the original representation (\ref{01}).
To see this,  we employ
the effective Hamiltonian approach worked out in Ref.\cite{sigrist} to treat the strong-coupling regime
of the Kondo-lattice model.
We start by rewriting the local lattice spin operators in the fermion-oscillator representation (\ref{fs}):
$\vec
S_i=\sum_{\sigma,\sigma'}f_{i\sigma}^{\dagger}\vec\tau_{\sigma\sigma'}f_{i\sigma'},$ where $f_{i\sigma}$
denotes the fermion operator subject to the on-site constraint,
$\sum_{\sigma}f_{i\sigma}^{\dagger}f_{i\sigma}=1.$

The new creation (annihilation) operators can then be introduced, \cite{sigrist}
$$\tilde c^{\dagger}_{i\sigma}=(1-n_{i}^d)f^{\dagger}_{i\sigma},\quad
\tilde c_{i\sigma}=(1-n_{i}^d)f_{i\sigma}$$
with $n_{i}^d=\sum_{\sigma}d^{\dagger}_{i\sigma}d_{i\sigma}.$ These operators are restricted to
$\tilde n^c_{i\sigma}\tilde n^c_{i-\sigma}=0$ for all sites, i.e., no double occupancy of $\tilde c$ states is allowed.
It is also clear that $$\tilde n^c_{i}=\sum_{\sigma}\tilde c^{\dagger}_{i\sigma}\tilde c_{i\sigma}=
(1-\tilde n^d_i)=(1-n^d_i).$$

In Ref.\cite{sigrist}, it is shown that the Kondo-lattice Hamiltonian
$$H=\sum_{ij\sigma} 2t_{ij}{d}_{i\sigma}^{\dagger} d_{j\sigma}
+\lambda
\sum_{i}\vec{S_i} \cdot \vec{M_i}$$ in the limit $\lambda\to\infty$
takes on the form
\begin{equation}
H=-\sum_{ij\sigma}t_{ij}\tilde c^{\dagger}_{i\sigma}\tilde c_{j\sigma}-\frac{3\lambda}{4}\sum_i\tilde n^d_i+{\cal O}
(t^2/\lambda).
\label{s1}\end{equation}
Notice now that the term $\propto 3\lambda/4$ in our representation (\ref{03}) exactly cancels out
the second term in Eq.(\ref{s1}). As for the spin exchange contribution $\propto J$ in (\ref{03}), it takes the form
\begin{equation}
J\sum_{ij}(\vec Q_i^c\vec Q_j^c-\frac{1}{4}\tilde n^c_{i}\tilde n^c_{j})
\label{s2}.\end{equation}
Collecting all this together, we get that, in the limit $\lambda\to\infty$, Eq.(\ref{03})
is equivalent to the original representation (\ref{01}).

Let us now demonstrate this equivalence rederiving
the ground-state energy of the $1d$ Hubbard model at $U=\infty$
in terms of the Kondo-type representation of the $t-J$ model as given by Eq.(\ref{04}) at $J=0$.
If this is the case our representation ({\ref{04}) is indeed in agreement with
the well known exact result.

The exact ground-state energy of the $U=\infty$ Hubbard Hamiltonian
\begin{equation}
H_{U=\infty}=-\sum_{ij\sigma} t_{ij} \tilde{c}_{i\sigma}^{\dagger}
\tilde{c}_{j\sigma},\quad \tilde{c}_{i\sigma}=c_{i\sigma}(1-n_{i,-\sigma})
\label{1}\end{equation}
takes in $1d$ the form \cite{ogata}
\begin{equation}
E^{U=\infty}_{gr}/N_{site}=-\frac{2t}{\pi}\sin(\pi x),
\label{2}\end{equation}
where $x=1-\sum_{\sigma}<\tilde{c}_{i\sigma}^{\dagger}
\tilde{c}_{i\sigma}>$ is the density of vacancies.
On the other hand, for $J\propto t^2/U=0$
Eq.(\ref{04}) reads
\begin{eqnarray}
H_{U=\infty}&=&\sum_{ij\sigma}
(2t_{ij}+\frac{3\lambda}{4}\delta_{ij}){d}_{i\sigma}^{\dagger} {d}_{j\sigma} +\lambda
\sum_{i}\vec{S_i} \vec{M_i},
\label{3}\end{eqnarray}
where $\lambda\to \infty$. This is the exact representation of the $U=\infty$ Hubbard model Hamiltonian.

If  Eq.(\ref{3}) is correct it must reproduce exactly Eq.(\ref{2}).
To show this, consider the $1D$ strong-coupling Kondo Hamiltonian
\begin{equation}
H^{Kondo}=-\sum_{ij\sigma} t_{ij}{c}_{i\sigma}^{\dagger}
{c}_{j\sigma}+\lambda\sum_i\vec S_i\vec s_i,\quad \lambda\to +\infty,
\label{4}\end{equation}
where now $c_{i\sigma}$ stands for a conduction electron operator,
and $\vec s_i$ denotes the conduction electron spin operator.
The ground-state energy is found to be \cite{sigrist}
\begin{eqnarray}
E^{Kondo}_{gr}/N_{site}
&=&\frac{t}{\pi}\sin(\pi x)-\frac{3}{4}\lambda x +{\cal O}(1/\lambda),
\label{5}\end{eqnarray}
Comparing now Eqs.(\ref{4}) and (\ref{3}) immediately gives for the ground-state energy of the
Hamiltonian (\ref{3})
\begin{equation}
E_{gr}/N_{site}=-\frac{2t}{\pi}\sin(\pi x)=E^{U=\infty}_{gr}/N_{site},
\label{6}\end{equation}
as desired.
Note once more that the $3\lambda/4$ term in Eq.(\ref{3})
plays an essential role in stabilizing the ground state energy in the limit
$\lambda\to +\infty.$

The ground state of the Hamiltonian (\ref{3}) is represented by the noninteracting {\it spinless} fermions \cite{ogata},
$$H_{gr}=-\sum_{ij} t_{ij} {c}_{i}^{\dagger}
{c}_{j}, \quad \{c^{\dagger}_i,c_j\}=\delta_{ij}.$$
This state
cannot be adiabatically connected to a state of weakly interacting dopons and lattice spins.

\section{Appendix B}

There is a formal analogy between the present formulation and that of
the $U=\infty$ Hubbard model. Consider the
Hamiltonian
\begin{equation}
H_{Hubb} = \sum_{ij\sigma} t_{ij} c_{i\sigma}^{\dagger}
c_{j\sigma} + U\sum_i n_{i\uparrow}n_{i\downarrow}.
\label{hub1}\end{equation}
In the case $U
\to \infty$,  the system is subject to the constraint
$n_{i\uparrow}+n_{i\downarrow} \le 1$. This constraint is equivalent
to $\Upsilon^{G}_i=n_{i\uparrow}n_{i\downarrow}=0$. In this way,
when $\Upsilon^{G}_i$ acts on the unphysical (doubly
occupied) states we have $\hat\Upsilon^{G}_i
|unphys\rangle_i=|unphys\rangle_i$. Therefore, $
P_i^{G}=1-n_{i\uparrow}n_{i\downarrow}$ is a projection operator
that eliminates the unphysical state at site $i$. The gauge
transformation generated by this constraint,
$$c_{i\downarrow}\to c_{i\downarrow}e^{i\theta_i n_{i\uparrow}}, \,
c_{i\uparrow}\to c_{i\uparrow}e^{i\theta_i n_{i\downarrow}},$$ leaves
the projected electron operators $\tilde
c_{i\sigma}=P_i^{G}c_{i\sigma}P_i^{G}=
c_{i\sigma}(1-n_{i-\sigma})$ intact. The global projection
operator is the well known Gutzwiller projector $P^{G}=
\Pi_i P_i^{G}$. We can then impose the constraint writing
$H_{Hubb} = P^{G} \sum_{ij\sigma} t_{ij} c_{i\sigma}^{\dagger}
c_{j\sigma}P^{G}$, which is equivalent to
\begin{equation}
H_{Hubb}
=\sum_{ij\sigma} t_{ij} \tilde{c}_{i\sigma}^{\dagger}
\tilde{c}_{j\sigma}.
\label{hub2}\end{equation}
This representations is equivalent to Eq.(\ref{hub1}) at $U\to +\infty$.
From this point of view, the Kondo coupling parameter $\lambda$ in the spin-dopon
representation of the $t-J$ model plays
the role of the Coulomb repulsion parameter $U$ in the Hubbard model at infinitely large $U$.

\section{Appendix C}

The explicit form of the effective hopping
amplitude $t^{eff}_{ij}(t,\chi)$ in Eq.(\ref{12.1})
is determined in Refs.\cite{ew,ewo} by equating matrix elements of a physical operator in the reduced Hilbert space
spanned by the basis vectors (the triplet states are discarded)
$$\sim \prod_i\tilde a_{i\sigma_i}^{\dagger}|\Psi_0\rangle$$
to matrix elements of a certain bilinear form of the canonical fermion operators $d_{i\sigma}$
in the Hilbert space
with the canonical basis $$\prod_{i}d_{i\sigma_i}^{\dagger}|0\rangle.$$
Since simply equating two operators acting in different Hilbert spaces (not isomorphic to each other) is in fact a
meaningless procedure, we provide below a more accurate treatment to
explicitly bring out the actual meaning of the conjecture made in Refs.\cite{ew,ewo}

To this end, we need the following representations that can be found in \cite{pfk}:
\begin{equation}
\tilde a_{i\downarrow}=\sqrt{2}P_i\tilde d_{i\downarrow}P_i, \quad
\tilde a_{i\uparrow}=-\sqrt{2}P_i\tilde d_{i\uparrow}P_i.
\label{c1}\end{equation}
Here $\tilde d_{i\sigma}=d_{i\sigma}(1-n^d_{i\bar\sigma})$ is the on-site Gutzwiller projected
dopon operator and the projection operator $P_i=1-\Upsilon_i^{sd}$ singles out the
subspace spanned by the local spin-$1/2$ states and the spin-dopon singlet state.

Let us now consider the matrix element \cite{ew}
\begin{equation}
\langle\Psi_0|\tilde a_{j\uparrow}H_t\tilde a_{i\uparrow}^{\dagger}|\Psi_0\rangle,
\label{c2}\end{equation}
where
$$H_t=\sum_{ij\sigma}t_{ij}d^{\dagger}_{i\sigma}d_{j\sigma}.$$
In view of Eqs.(\ref{c1}), this can be rewritten as $(P\equiv \prod_iP_i)$
\begin{equation}
\sim \langle\Psi_0|P\tilde d_{j\uparrow}PH_tP\tilde d_{i\uparrow}^{\dagger}P|\Psi_0\rangle.
\label{c3}\end{equation}
At low doping concentration $x\ll 1$, one can drop the "tilde" sign over the dopon operators,
which brings the matrix element to the form  $(P|\Psi_0\rangle=|\Psi_0\rangle)$
\begin{equation}
\sim \langle\Psi_0|d_{j\uparrow}\tilde H_td_{i\uparrow}^{\dagger}|\Psi_0\rangle,
\label{c4}\end{equation}
where
$$\tilde H_t\equiv PH_tP=\sum_{ij\sigma}t_{ij}\tilde a^{\dagger}_{i\sigma}\tilde a_{j\sigma}.$$

The key approximation made in Refs.\cite{ew,ewo}
amounts then to discarding the $P$ projection accompanied by a simultaneous
renormalization of the hopping amplitude:
$$\tilde a_{i\sigma}=sign(\sigma)\sqrt{2}P_id_{i\sigma}P_i\to sign(\sigma)\sqrt{2}d_{i\sigma},
\quad t_{ij}\to t^{eff}_{ij},$$
which yields for the matrix element
\begin{equation}
\sim \langle\Psi_0|d_{j\uparrow}H_{t_{eff}}(d^{\dagger},d)d_{i\uparrow}^{\dagger}|\Psi_0\rangle,
\label{c5}\end{equation}
To explicitly fix $t^{eff}_{ij}$,
the matrix elements (\ref{c2}) and (\ref{c5}) are then equated
to each other.
It should be stressed  that while the Gutzwiller projection for the dopon operators can indeed be safely
discarded at low doping, this is obviously not the case for the NDO projection operator $P$.


\begin{thebibliography}{99}
\bibitem{w3} In \cite{wen2}, the underdoped cuprates is described as a "Luttinger-volume violating
liquid" (LvvFL) with a spin liquid of fermionic spinons. The LvvFL state is qualitatively the same
as the FL$^*$ state.
\bibitem{zero} For infinite $\lambda$, the Kondo temperature, which is essentially the singlet-triplet
separation, is also infinite. The exponential variation $T_K=De^{-1/\rho \lambda}$
is only valid for small $\lambda$.\cite{lacroix}
\end{thebibliography}
\end{document}